\documentclass[11pt, a4paper, logo, copyright, nonumbering]{googledeepmind} %

\usepackage[authoryear, sort&compress, round]{natbib}
\title{(Unfair) Norms in Fairness Research:\\A Meta-Analysis}
\shorttitle{(Unfair) Norms in Fairness Research: A Meta-Analysis}

\correspondingauthor{jjchien@ucsd.edu}


\usepackage[english]{babel}
\usepackage{amsthm}

\usepackage{array,booktabs,tabularx,multirow, multicol,colortbl,hhline,makecell}

\usepackage{graphicx,xcolor,caption,placeins,wrapfig}

\usepackage[utf8]{inputenc}
\usepackage{graphicx}
\usepackage{makecell}
\usepackage{caption}
\usepackage{subcaption}

\usepackage{array}
\usepackage{makecell}
\usepackage{orcidlink}

\author[1]{Jennifer Chien~\orcidlink{0009-0009-8768-1761}}
\author[2]{A. Stevie Bergman~\orcidlink{0000-0002-4331-1357}}
\author[2]{Kevin R. McKee~\orcidlink{0000-0002-4412-1686}}
\author[2]{Nenad Tomasev~\orcidlink{0000-0003-1624-0220}}
\author[3]{Vinodkumar~Prabhakaran~\orcidlink{0000-0003-3329-2305}}
\author[3]{Rida Qadri~\orcidlink{0000-0001-5690-0997}}
\author[2]{Nahema Marchal~\orcidlink{0000-0002-8518-3840}}
\author[2]{William Isaac~\orcidlink{0000-0002-1297-5409}}

\affil[1]{University of California San Diego}
\affil[2]{Google DeepMind}
\affil[3]{Google Research}

\begin{document}

\begin{abstract}
Algorithmic fairness has emerged as a critical concern in artificial intelligence (AI) research. However, the development of fair AI systems is not an objective process. Fairness is an inherently subjective concept, shaped by the values, experiences, and identities of those involved in research and development. To better understand the norms and values embedded in current fairness research, we conduct a meta-analysis of algorithmic fairness papers from two leading conferences on AI fairness and ethics, AIES and FAccT, covering a final sample of 139 papers over the period from 2018 to 2022. Our investigation reveals two concerning trends: first, a US-centric perspective dominates throughout fairness research; and second, fairness studies exhibit a widespread reliance on binary codifications of human identity (e.g., ``Black/White'', ``male/female''). These findings highlight how current research often overlooks the complexities of identity and lived experiences, ultimately failing to represent diverse global contexts when defining algorithmic bias and fairness. We discuss the limitations of these research design choices and offer recommendations for fostering more inclusive and representative approaches to fairness in AI systems, urging a paradigm shift that embraces nuanced, global understandings of human identity and values.
\end{abstract}

\maketitle

\section{Introduction}
\label{sec:intro}
The widespread adoption of artificial intelligence (AI) brings with it the potential for substantial harm. AI systems frequently encode and amplify historical biases~\citep{lum2916, buolamwini2018gender, eubanks2018automating, dignazio&klein2020data, zou2023ai}, thus exacerbating and perpetuating discrimination against marginalized communities~\citep{eubanks2018automating, benjamin2020race, irani2010postcolonial}. As a consequence, \textit{algorithmic fairness} has emerged as a growing priority for AI developers, ethicists, policymakers, and regulators~\citep{pfeiffer2023algorithmic, kleanthous2022perception, lepri2018fair, zarsky2016trouble}.

There is no single definition of fairness~\citep{mehrabi2021survey,barocas2023fairness}. Accordingly, approaches to algorithmic fairness vary widely, encompassing individual and group perspectives, parity and equity considerations, and beyond~\citep{dwork2012fairness, binns2020apparent, chien2023fairness}. While this pluralism has allowed researchers to generate a diverse toolkit of metrics, techniques, and frameworks for mitigating bias, it also introduces a crucial challenge: the inherent subjectivity of fairness itself.

Fairness is not a universal, abstract concept. It is deeply intertwined with the values, experiences, and identities of those involved in the research process~\citep{guillemin2004ethics, fook1999reflexivity, mccabe2009reflexivity, bunge2018critical, iliadis2016critical}.
Thus, fairness research is sensitive to contextual questions of who is doing the work, what they are studying, and how they are studying it. A comprehensive understanding of fairness necessitates a conscious effort to acknowledge and understand the values embedded in the research process itself.

In this work, we undertake a reflexive meta-analysis of the research literature on algorithmic fairness, exploring the values and perspectives embedded within the research process itself. We focus our meta-analysis on the AAAI/ACM Conference on Artificial Intelligence, Ethics, and Society (AIES) and the ACM Conference of Fairness, Accountability, and Transparency (FAccT), two prominent conferences on algorithmic fairness. Our investigation reveals two concerning trends: a US-centric bias in authorship and definitions of sensitive attributes, and the predominance of binary formulations for sensitive attributes (e.g., ``Black/White'' or ``male/female''). These findings indicate that predominant approaches to fairness radically oversimplify the complexities of identity and fail to reflect the diversity of lived experiences with AI across the globe. Overall, the results from our meta-analysis lead us to join the growing chorus of calls for a paradigm shift in fairness research. Moving forward, the research community must embrace a more nuanced and contextual understanding of human values, supporting the development of genuinely representative, inclusive, and fair AI systems.

\section{Related Work}\label{sec:related work}

Our meta-analysis builds upon a growing body of research surveying and critiquing the fields of AI ethics and algorithmic fairness.

\subsection{Data and Geographic Bias}
In recent years, critical researchers have increasingly examined AI ethics and fairness research for geographic bias. Close scrutiny reveals that the datasets used within these fields often fail to accurately represent the global population. \citet{septiandri2023weird} survey FAccT proceedings and finds a bias towards Western, educated, industrialized, rich, and democratic perspectives participants and data. \citet{abdu2023empirical} examine racial categories in FAccT proceedings through the lens of institutional influences and values, finding that projects adopt racial categories inconsistently, often following country-specific legal frameworks, while rarely explicitly describing or justifying their choices. Similarly, \citet{koch2021reduced} touch on the consequences of geographic bias in their exploration of the increasing, over-concentrated usage of a limited group of datasets in machine learning research.

\subsection{Western-Centric Values and Perspectives}
AI ethics research often exhibits a bias toward countries in the Global North, presenting their cultural perspectives and values as universal and globally fungible. \citet{hagerty2019global} study this phenomenon in global social science scholarship and argue that mitigating the worldwide harms caused by AI deployments in the Global North will require deep understandings of a broad set of geographical, cultural, and social contexts. In an examination of the proceedings at FAccT and the ACM CHI Conference on Human Factors in Computing Systems, \citet{van2023methodology} document a bias toward the US across study design, participant recruitment, and country affiliation of paper authors. \citet{birhane2022forgotten} critique ethics papers at AIES and FAccT as relying on speculative, theoretical foundations from Western philosophy, resulting in a dearth of concrete use cases and a lack of acknowledgement for afflicted communities. \citet{sambasivan2021re} similarly examine the philosophical roots, legal frameworks, and axes of discrimination within AI ethics to refocus on what values and norms translate and fail to translate to applications in India. Motivated by the non-portability of algorithmic fairness to India, \citet{sambasivan2021seeing} reviews data practices in the machine learning pipeline and discusses the need for expanded representation, calling for the creation and maintenance of work focused on the Global South.

\section{Methods}
\label{sec:methods}

This study aims to examine the practices of problem selection and formulation within algorithmic fairness research. What values do researchers explicitly and implicitly elevate? How does each research project approach its object(s) of study? How does each project interpret and present its results?

In answering these questions, we aim to reflexively surface and record common practices within fairness research—and subsequently to inform future best practice. We employed a meta-analysis methodology for our study, given its ability to synthesize and critically analyze existing research~\citep{borenstein2021introduction}.

\begin{table}[htbp]
\centering
\begin{tabular}{p{0.3\linewidth}p{0.65\linewidth}}
  \toprule
  \textbf{Collected Label} & \textbf{Definition}\\
  \toprule
  Paper Title & As provided \\ 
  Author Country Affiliation & IBAN two-letter country codes for each author, comma separated\\
  Year Published & [\textit{2018}; \textit{2019}; \textit{2020}; \textit{2021}; or \textit{2022}]\\
  Venue & [\textit{AIES} or \textit{FAccT}]\\
  Study Type & All that apply from: [\textit{Retrospective} (uses existing datasets); \textit{Theory} (develops novel definitions, proofs, or guarantees); \textit{Prospective} (collects new datasets)]\\
  \midrule
  
  Dataset Name & As provided \\
  Dataset Origin Country & [\textit{Yes}; \textit{No}; or \textit{Unspecified}]\\
  Data Type & [\textit{Tabular}; \textit{Text}; \textit{Time Series}; \textit{Still Images}; \textit{Video}; or \textit{Audio}]\\
  Topic Domain & [\textit{Criminal Justice}; \textit{Education}, \textit{Media} (e.g., entertainment, news); \textit{Finance}; \textit{Health/Medicine}; \textit{Hiring}; \textit{Social Media} (e.g., networks); \textit{Informational} (e.g., Wikipedia); \textit{Social Welfare}; or \textit{Obscured}]\\
  Sensitive Attribute Category(s) Studied  & As specified (e.g. ``Race'', ``Geolocation'')\\
  Attribute Specific Labels & If a sensitive attribute was specified, the labels used (e.g. ``Asian'', ``Female'', ``Age (1-18)'')\\
  Degree of Sensitive Attribute Measurement & [\textit{No Acknowledgement}; \textit{Acknowledgement}; or \textit{Acknowledgement and Mitigation}]\\
  Degree of Sensitive Attribute Limitation Addressed & [\textit{No Acknowledgement}; \textit{Acknowledgement}; or \textit{Acknowledgement and Mitigation}]\\
  Proxy used for Sensitive Attribute & [\textit{Yes}; \textit{No}; or \textit{Not Applicable (NA)}]\\
  Proxy Category & If a proxy is used, the category the proxy is used for (e.g. ``Race'')\\
  Proxy Label & If a proxy is used, the labels used (e.g. ``Skin Tone'')\\
  Does the paper address intersectionality? & [\textit{Yes}; \textit{No}; or \textit{NA}]\\
  Intersectional Category(s) & If intersectional, the sensitive attribute category(s) studied (e.g. ``Race'')\\
  Degree of Dataset Limitation & [\textit{No Acknowledgement}; \textit{Acknowledgement}; or \textit{Acknowledgement and Mitigation}]\\
  \midrule
  
  Fairness Metric Name & As provided \\
  Metric Type & All that apply from: [\textit{Group}; \textit{Individual}, \textit{Counterfactual}; or \textit{Contrastive}]\\
  Explicit Inputs & Whether the sensitive attributes are explicit inputs to the model [\textit{Yes}; \textit{No}; or \textit{NA}]\\
  Performance Trade-off & [\textit{Yes}; \textit{No}; or \textit{Unclear}]\\
  Deep Learning & Whether the model uses deep learning [\textit{Yes}; \textit{No}; or \textit{NA}]\\
  Participatory & Whether the paper takes a participatory approach [\textit{Yes}; \textit{No}; or \textit{NA}]\\
  Human Factors & Degree to which the paper addresses how humans factor into the decision-making process [\textit{No Acknowledgement}; \textit{Acknowledgement}; or \textit{Acknowledgement and Mitigation}]\\
  Pipeline Intervention & The stage of the pipeline where the intervention is proposed [\textit{Pre-processing}; \textit{In-processing}; \textit{Post-processing}; or \textit{Multiple Points}]\\
  \bottomrule
\end{tabular}
\caption{Table of Collected Labels \& Definitions. We pre-generated initial levels for variables such as ``Data Type'' and expanded levels for instances that did not fit the existing scheme.}
\label{Fig::Data Table}
\end{table}

\subsection{Research Questions}

To contribute to ongoing efforts to identify the normative choices and biases within fairness research, we investigate the following research questions:

\begin{enumerate}[align=left,labelwidth=\parindent,labelsep=5pt,leftmargin=*,start=1,label={\bfseries RQ\arabic{enumi}:}]
    \item What geographic and cultural biases are present in algorithmic fairness research?
    \item What are the implicit norms embedded in the research process for algorithmic fairness, especially those affecting the diversity and representativeness of the research?
\end{enumerate}

\subsection{Data and Analysis}

Our review investigated papers from two flagship conferences on algorithmic fairness and AI ethics: the AAAI/ACM Conference on Artificial Intelligence, Ethics, and Society (AIES) and the ACM Conference of Fairness, Accountability, and Transparency (FAccT). Both have grown substantially since their establishment in 2018, with interdisciplinary proceedings (including computer science, law and policy, social sciences, ethics and philosophy) and attendees (including researchers, policymakers, and practitioners). Together, the conferences represent principal sites of the research discourse on algorithmic fairness~\citep{acuna2021ai, birhane2022forgotten}. We collected all papers included in the conferences' proceedings from 2018 (their inception) through 2022 (the most recent year available during our data collection period), accessing the papers through the conference websites and the ACM Digital Library. This initial sample comprised 265 and 416 papers from AIES and FAccT, respectively. We subsequently removed non-archival and abstract-only papers, and then filtered our sample to papers that specify at least one formal fairness definition and one model-based decision-making process. This final sample included 139 papers (52 from AIES and 87 from FAccT, respectively).

We focus our annotation and analysis on several crucial aspects of the sampled research, including dataset characteristics, author information, and modeling choices. To ensure consistency in our data collection, we collaboratively developed and defined our coding scheme. During this process, we independently coded two papers and compared results to clarify any points of confusion or disagreement.

We begin by examining the datasets employed in each paper, recording the dataset name, type (e.g., tabular, text), and topic domain. We further document the sensitive attributes studied, noting the specific categorization scheme applied to attribute labels, label definitions, and collection methods. We pay particular attention to whether papers employed proxies to study sensitive attributes, whether the papers discussed intersectionality, and the extent to which papers acknowledged and mitigated limitations related to these design choices. For simplicity, we categorize the degree of these practices on a three-point scale: (\textit{no acknowledgement}, \textit{acknowledgement}, and \textit{acknowledgement and mitigation}). Additionally, we note any available information on dataset provenance, allowing for identification of potential biases towards specific regions or populations and for comparisons with authors' country affiliations.

We next catalogue author information. We document the countries noted in author affiliations and record whether each paper incorporated participatory feedback throughout the development process. We also assess the extent to which authors acknowledged any interaction effects of human-in-the-loop processes in the decision-making process, employing the same three-point scale as for acknowledging dataset limitations.

We subsequently examine model design decisions. We note whether each model explicitly received sensitive attributes as inputs, record the name assigned to the model's fairness metric, and classify the metric within \textit{contrastive}, \textit{counterfactual}, \textit{group}, or \textit{individual fairness}. We also record any mitigation methods used, categorizing each as a \textit{constraint} or \textit{objective}.

After coding the full sample, we quantify temporal and aggregate trends in these metrics. Guided by our research questions, we pay particular attention to author country affiliation, dataset country affiliation, sensitive attribute labels (number and instantiation), and label definition.

Finally, we enrich our empirical findings through a \textit{sociotechnical} analysis. Sociotechnical approaches emphasize the inherently intertwined nature of social and technical systems, allowing them to explore the ways in which social dynamics shape technological development~\citep{cherns1976principles, dolata2022sociotechnical}. By applying this framework, we gain a more holistic understanding of the complex interplay between technological systems, human actors, and their broader societal context within algorithmic fairness research. We also adopt a reflexive stance, acknowledging that our own positions as researchers inevitably shape our perspective and analysis~\citep{collins1992black, harding1991whose, haraway1988situated}. This self-awareness helps us to identify implicit norms and assumptions embedded within the field, leading to a more nuanced understanding of the practice of algorithmic fairness research.

\begin{figure}[ht!]
\centering
\includegraphics[width=0.90\linewidth]{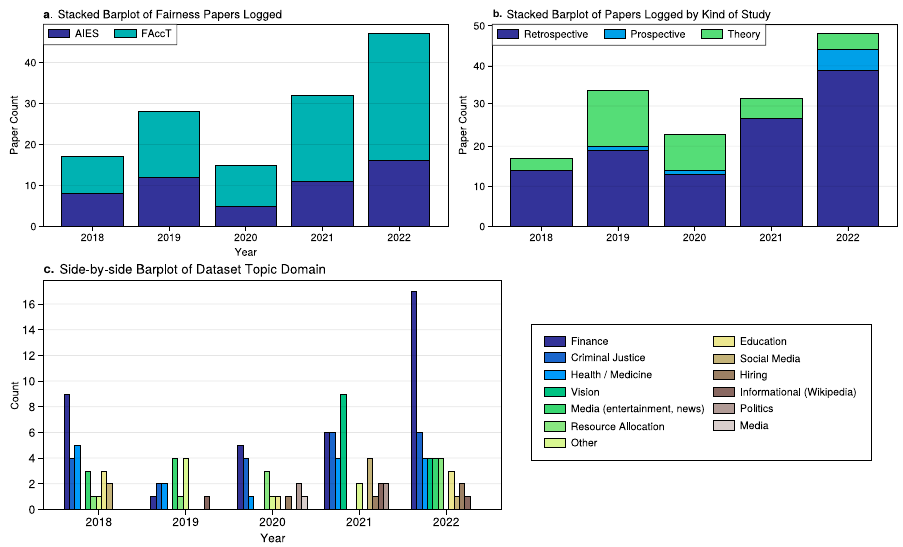}
\caption{Conference, Study Type, and Dataset Topic Domains Over Time. (a) The number of papers published by each conference trends upward over time. (b) Over all years, a majority of papers are retrospective (conducting empirical analyses on pre-existing datasets). (c) Papers examine an increasing variety of topic domains over time. Overall, finance generally prevails as the most popular dataset domain application, followed by criminal justice.}
\label{Fig::Papers Logged and Study Type}
\end{figure}

\section{Results}\label{sec:findings}

\subsection{Quantitative Findings}

Overall, we examine a total of 139 papers, with 52 and 87 from AIES and FAccT, respectively (\hyperref[Fig::Papers Logged and Study Type]{Figure~\ref{Fig::Papers Logged and Study Type}}a). These papers map to a total of 124 unique datasets. %
Retrospective studies (research empirically examining fairness within pre-existing datasets) represented the majority of papers, followed by theory (research contributing theoretical guarantees of fairness definitions). The prevalence of prospective studies (research collecting and empirically examining novel data) remained relatively low across most years, increasing the most in 2022 (\hyperref[Fig::Papers Logged and Study Type]{Figure~\ref{Fig::Papers Logged and Study Type}}b). Dataset domains spanned a wide range of topics, including finance, criminal justice, health, and politics. Finance emerged as the majority domain within most years (\hyperref[Fig::Papers Logged and Study Type]{Figure~\ref{Fig::Papers Logged and Study Type}}c).
\hyperref[App::Study Design]{Appendix: Study Design} and \hyperref[App::Fairness Definitions]{Appendix: Fairness Definitions} provide additional results concerning fairness formulations.

\subsubsection{Data and Authorship Provenance}
\begin{figure}[htbp]
\centering
\includegraphics[width=0.95\textwidth]{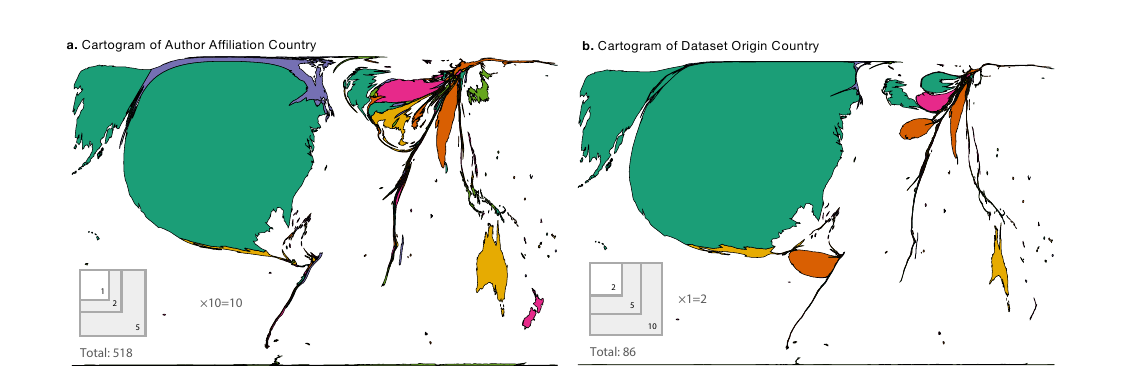}
\caption{Distribution of Author Country-Affiliation and Dataset Country-Affiliation.
These cartograms represent country sizes proportional to the count of (a) author affiliations and (b) datasets attributed to each country. The US emerges as the most highly represented country for both authorship origin and data provenance. Graphics created with \citet{gastner2018fast}. 
}
\label{Fig::Cartogram}
\end{figure}

\hyperref[Fig::Cartogram]{Figure~\ref{Fig::Cartogram}} visualizes geographic patterns in authorship provenance and dataset provenance through cartograms. These depict country size proportional to their representation. 
In terms of author affiliation, the US held the greatest count and proportion of affiliations (412, 78.0\%), followed by Germany (19, 3.5\%), Canada (14, 2.7\%), the United Kingdom (13, 2.5\%), Australia (12, 2.3\%), Italy (11, 2.0\%), Switzerland (10, 1.9\%), and India (10, 1.9\%). 
When broken down by year, we find that 80.6\% of the papers published each year feature at least one author with ties to the United States (see e.g. \hyperref[App::Authorship Affiliation By Year]{Appendix: Data and Authorship Provenance}). For dataset provenance, the US again emerged as the most common country of origin (77, 72.6\%), followed by Germany (9, 8.5\%) and then Colombia, Rwanda, Australia, Burundi, Costa Rica, Finland, Hungary, Iceland, India, Kenya, Malawi, Mexico, Mozambique, New Zealand, Portugal, Senegal, South Africa, Sweden, Switzerland, Tanzania, Thailand, Uganda, the United Kingdom, Zambia, and Zimbabwe, all with fewer than three datasets ($\leq$2.8\% each, 18.9\% collectively). In our sample, 15.9\% of papers acknowledge at least one dataset limitation. In addition, 1.4\% employ some form of mitigation, such as introducing additional data \citep[e.g.,][]{dixon2018measuring} or re-annotating the data with an expert \citep[e.g.,][]{buolamwini2018gender}.

\begin{figure}[htbp]
\centering
\includegraphics[width=0.7\textwidth]{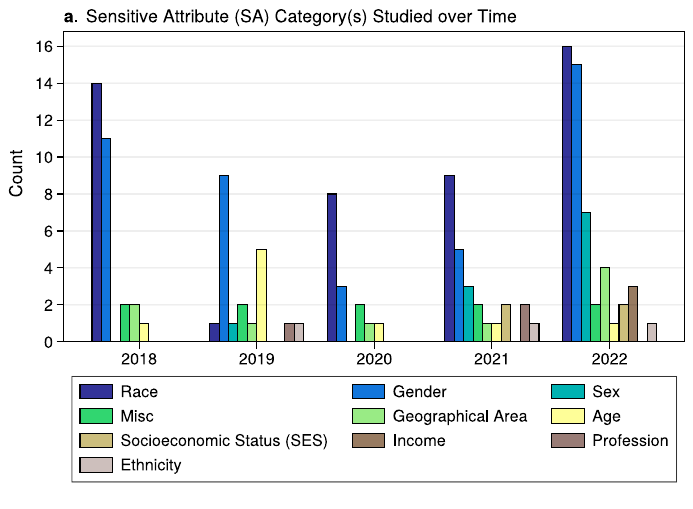}
\caption{Sensitive Attributes Studied Over Time. Count reflects the number of times papers in our sample analyzed each sensitive attribute category in a given year.}
\label{Fig::SAtop10}
\end{figure}

\subsubsection{Sensitive Attributes}
To account for the wide range of variables studied by the algorithmic fairness community, we define ``sensitive attributes'' as the features across which a study aims to measure or ensure fairness. This broad definition includes features protected by law (e.g., ``protected attributes'' such as race or gender) as well as those with potential for social or economic discrimination (e.g., income). Papers in our sample examined a total of 49 sensitive attributes, in which gender, race, and age emerged as the top three categories of sensitive attributes in our sample (\hyperref[Fig::SAtop10]{Figure~\ref{Fig::SAtop10}}).\footnote{
Some papers used multiple categories interchangeably (e.g., ``sex/gender'', ``race/ethnicity''), without providing definitions for each constituent category. 
While we believe that many of these categories represent distinct social constructs (e.g.,~\citealp{unger1979toward, lips2020sex}), unpacking these groupings retrospectively proved impractical given the available information.
Thus, for the purposes of this work, we combine categories such as sex and gender.
However, we urge future work to carefully consider and define the specific social constructs under investigation.}

\begin{figure}[htbp]
\centering
\includegraphics[width=0.9\textwidth]{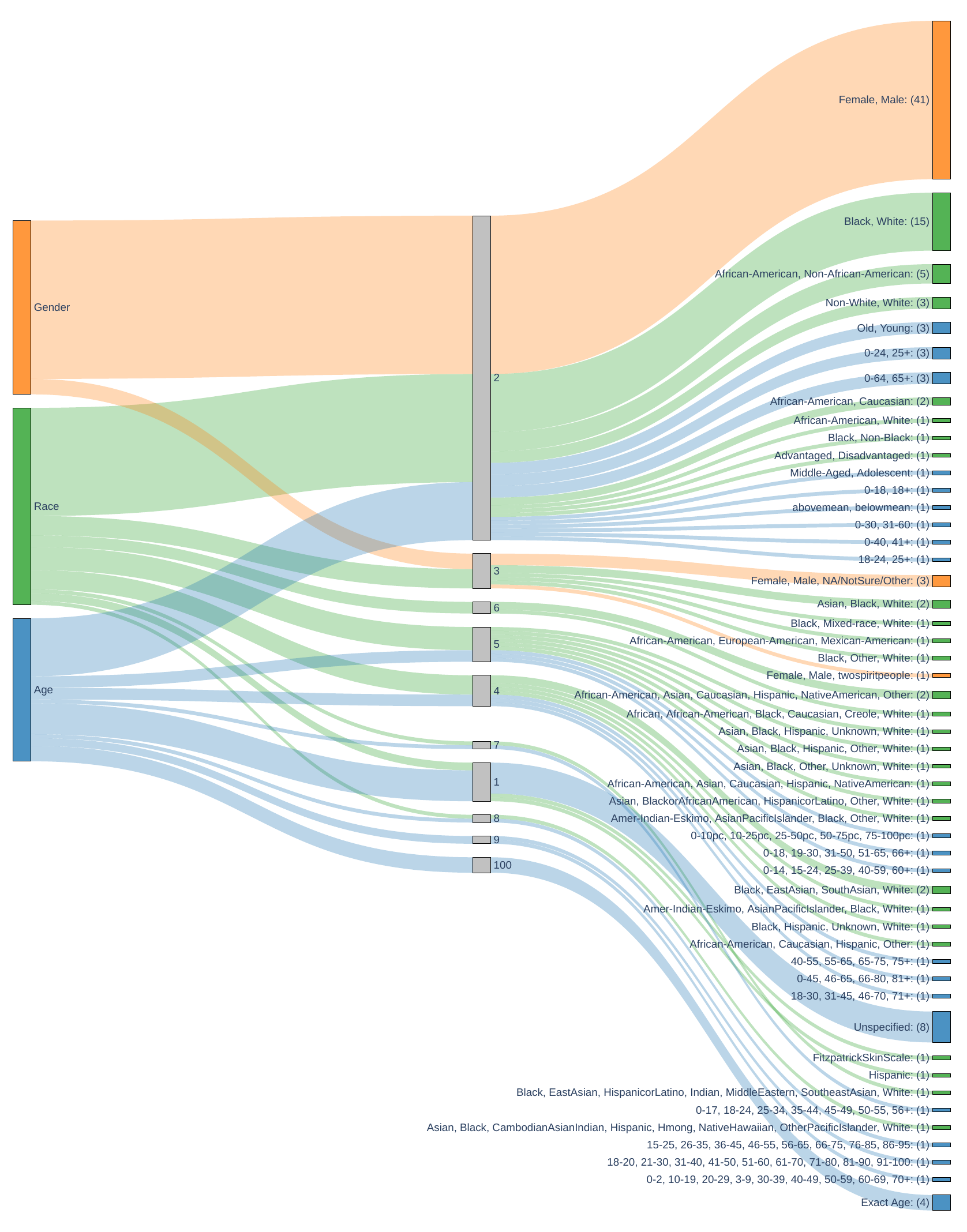}
\caption{Distribution of Label Formulations for Sensitive Attributes Across Fairness Studies. This alluvial diagram illustrates the range of formulations that fairness studies use to label sensitive attributes. Each colored band represents, on the left, one of the three most frequently studied sensitive attributes in our sample, and on the right, the number of studies utilizing a particular formulation. The grey bars in the middle indicate the number of unique categories employed within each formulation. Studies predominantly formulate gender as a ``female/male'' binary. Across race, the top two formulations are ``Black/White'' and ``African-American/non-African-American''. Finally, age shows some increased diversity in the number of categories considered, though papers still exhibit a strong tendency towards binary formulations (e.g., 0-24 and 25+ or 0-64 and 65+).
}
\label{Fig::Alluvial Diagram All}
\end{figure}

\begin{figure}[htbp]
\centering
\includegraphics[width=0.5\textwidth]{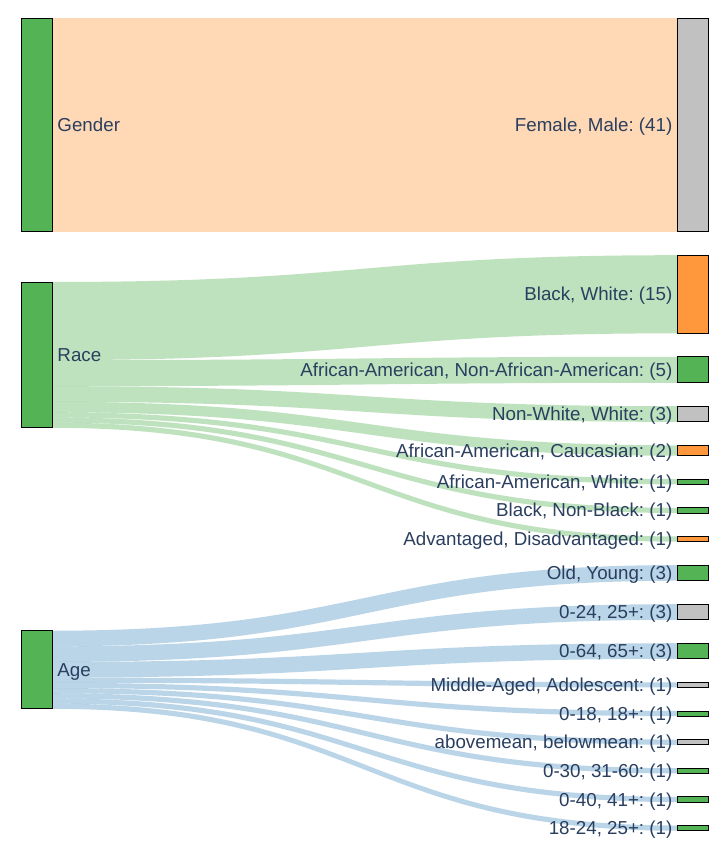}
\caption{Binary Formulations for Sensitive Attributes. This alluvial diagram highlights the binary label formulations used for the three most studied sensitive attributes (gender, race, and age) within fairness studies. Each colored band represents, on the left, one of the three most frequently studied sensitive attributes in our sample, and on the right, the number of studies utilizing a particular formulation. The grey bars in the middle indicate the number of unique categories employed within each formulation.}
\label{Fig::Alluvial Diagram Binary}
\end{figure}

Our analysis revealed substantial variation in the labels assigned to each sensitive attribute (that is, variation in the category options for sensitive attributes, such as ``female'' or ``Black''), even when papers relied on the same dataset (see Figures~\hyperref[Fig::Alluvial Diagram All]{\ref{Fig::Alluvial Diagram All}} \&~\hyperref[Fig::Alluvial Diagram Binary]{\ref{Fig::Alluvial Diagram Binary}}). Race labels included ``Black/African American'', ``White/Caucasian'', ``Asian/South Asian/East Asian/Pacific Islander'', ``Hispanic/Latino/Mexican-American'', ``Native American/American Indian/Eskimo'', and ``unknown/other''. Age labels exhibited similar inconsistencies, employing various cutoff points such as 0-17/18+, 0-24/25+, 0-64/65+, as well as broader categorizations like ``young/old'' and systems using intervals of 5, 10, or 15 years. Gender displayed the least variation, with most papers employing ``female/male''. Although some datasets included third categories such as ``NA/other/not sure''~\citep{ekstrand2018all, yang2022enhancing, usunier2022fast}, ``non-binary or choose not to disclose ''~\citep{schoeffer2022there}, or ``two spirit''~\citep{suresh2022towards}, papers generally excluded these categories from analyses and empirical demonstrations. 

Most papers (89.7\%) failed to provide any definition for sensitive attribute labels. This issue was particularly pronounced for age, with 20.6\% of studies investigating age neglecting to define the age ranges that they used.
Beyond simply defining the labels, most papers also omitted information regarding the source of their labels; researchers rarely clarified whether the labels originated from third-party observation or self-disclosure. Similarly, papers often did not provide any additional data or criteria that might have been used to determine the labels. 
Papers in our sample, for instance, frequently aggregated multiple categories into broader labels---such as ``young'' and ``old'' or ``African-American'' and ``non-African American''---without providing any explanation.

\subsection{Sociotechnical Analysis}

To further contextualize our quantitative findings, we conduct a sociotechnical analysis to explore and make explicit the unstated influences shaping algorithmic fairness research. This analysis focuses on identifying potential geographical, cultural, and implicit norms within the fairness research community and exploring how these biases might affect the diversity and representativeness of conducted studies. Given the interconnected nature of social and technical systems~\citep{cherns1976principles, dolata2022sociotechnical}, these social norms and biases will ultimately influence the technological outcomes of algorithmic fairness research. Consequently, we consider whether these norms and biases affected the scope of research, the disclosure of limitations, and assumptions about global fungibility. For this analysis, we drew on our own experiences as scientists studying algorithmic fairness and as members of marginalized communities affected by deployed systems~\citep{collins1992black, haraway1988situated}.

Despite the existing discourse on bias and limitations in fairness research (see \hyperref[sec:related work]{Related Work}), we observed two striking patterns across the empirical proceedings at AIES and FAccT: a disproportionate bias toward the US in both author affiliations and dataset origin, and a pervasive tendency towards binary formulations of sensitive attributes, particularly for gender, race, and age.

\subsubsection{US-Centrism and the Limits of Global Fungibility}
\label{Subsec::US-Centrism}

Our meta-analysis reveals a clear and substantial bias toward the US in algorithmic fairness research. We find that 80.6\% of papers published each year feature at least one US-based author, and that the US also dominates dataset provenance (72.6\%; see \hyperref[Fig::Cartogram]{Figure~\ref{Fig::Cartogram}} \& \hyperref[App::Authorship Affiliation By Year]{Appendix: Data and Authorship Provenance}). In our sample, papers rarely acknowledged or addressed potential biases arising from the predominance of the US in author affiliations and dataset origins.\footnote{We observe similar patterns among other research hubs in the Global North. For instance, Germany (representing 3.5\% of author affiliations and 8.5\% of dataset origins), the UK (2.5\% and 2.8\%), and Australia (2.3\% and 2.8\%)---though less overall prominent than the US---each fail to acknowledge or mitigate their potential geographic biases.}

The very labels that researchers use to define sensitive attributes reveal a substantial bias toward American norms and values. For instance, papers in our sample predominantly frame race as a Black/White binary (e.g., ``White/Non-White'', ``Black/Non-Black''). This simplification not only obscures nuances within Black identity (e.g., distinguishing between African and African-American experiences;~\citealp{agyemang2005negro}), but also homogenizes a diverse array of backgrounds under the label ``White'' (e.g., Mexican, Caribbean, Puerto Rican, Middle Eastern, and North African identities; \citealp{overmyer2013good, denton1989racial, landale2002white, awad2021identity}). This binary framing reflects a cultural phenomenon specific to the US: in particular, the historical and ongoing salience of the Black/White racial dichotomy, within American culture~\citep{goldstein2006price, jones2015black}. Similarly, a number of papers and datasets exploring fairness with respect to age use 25 and 65 years as key demarcation points (Figures~\ref{Fig::Alluvial Diagram All} \&~\ref{Fig::Alluvial Diagram Binary}). This practice echoes US legal frameworks regarding adulthood, dependence status, and retirement~\citep{hamilton2016adulthood, agich2003dependence, dailey2018new, boni2022legal}.

These specific demarcation points cohere with US societal structures, but do not universally translate to other countries with different legal and social norms surrounding adulthood, retirement, and other life stages~\citep{malone2011global, bhabha2008independent, oecd2021pensions, oecd2009social}. These choices coalesce into a clear pattern of \textit{US-centrism}---the tendency to view the world through the political, economic, and social lens of American society \citep{shabbar2017authoritative, chaturvedi2007whose}. 

The US-centrism pervading algorithmic fairness research inevitably shapes the field's outputs. Research does not exist in a vacuum---it is profoundly influenced by the socio-cultural and legal environments in which it originates~\citep{hagerty2019global, cole2017science, turner2023legal}. The definition of demographic categories demands particular attention here. Though often framed as objective and neutral, demographic labels carry a long history of weaponization, particularly against marginalized groups. Notions of lineage, morality, aesthetics, sexuality, and gender have long been used to define and differentiate social groups within the US, often to the detriment of those deemed ``different'' (see e.g.~\citealp{hollinger2005one, kitano1984asian, decuir2014proving, harris1993whiteness, gilman1999making, said1993culture, tasca2012women, kamin1998race, gooren2015medicalization}). By neglecting to explicitly define sensitive attributes and acknowledge this history of US-centric categorization, algorithmic fairness research obscures the power dynamics inherent in its label choices. The lack of precise definitions effectively shields these conceptual frameworks from the critical scrutiny needed to deconstruct and remake systems of power. As a result, algorithmic fairness research risks encoding US-centric inequalities and social hierarchies into the very systems intended to promote fairness.

The uncritical adoption of US-centric norms extends beyond label choices to additional aspects of research design. For instance, researchers in our sample rarely justified their choice of fairness metrics, evaluation datasets, or even problem framing. These design choices, often shaped by the US socio-political context, are presented as default or universally applicable without acknowledging their inherent limitations. This assumption of global fungibility obscures the need for localized benchmarks and diverse perspectives (cf. \citealp{castro2022north}). The subsequent lack of effort to contextualize research further entrenches the predominance of US-specific norms and design choices. As a result, the assumed generalizability of these research findings to other contexts goes largely unquestioned, and the perceived preeminence of US-centric values in the field remains unchallenged~\citep{birhane2022forgotten}.

\subsubsection{Binary Formulations and the Erasure of Intersectional Realities}
\label{Subsec::Binary Formulations}

Our meta-analysis reveals a second concerning trend in algorithmic fairness research: the pervasive use of \textit{binary formulations} for sensitive attributes (see \hyperref[Fig::Alluvial Diagram All]{Figure~\ref{Fig::Alluvial Diagram All}}). This binary framing imposes an overly simplistic structure that obscures the complexities of lived experiences and positions groups as isolated, opposing poles on a single axis of power and privilege. For instance, fairness studies often categorize age as ``young'' versus ``old'', despite the enormous range of communities that exist within these two age groups. This dualistic logic presumes homogeneity within each class, ignoring the diverse experiences and identities within those broad categories.

These assumptions carry profound implications for algorithmic fairness.
Binary formulations minimize the unique challenges faced by members of understudied groups. For instance, the ``male/female'' binary erases individuals who identify outside this dichotomy. Similarly, studies that focus on ``Black/White'' comparisons in the US obscure the experiences of Hispanic, Asian, and Indigenous communities. Yet research efforts that adopt a binary approach often incorrectly assume that their findings concerning harms and mitigations will translate seamlessly across groups.

Ironically, by treating each of its categories in isolation, this framework equates distinct experiences of discrimination, assuming that---for instance---bias toward racial and gender minorities pose interchangeable challenges.
In essence, the binary approach to fairness encourages a view of discrimination as a one-size-fits-all problem, neglecting the nuanced ways in which different forms of oppression intersect and interact. By overlooking the interconnected nature of social categories, binary formulations fail to address the compounding effects of intersectionality in the matrix of domination: axes of discrimination are neither separable nor additive~\citep{crenshaw1989demarginalizing,collins1992black}.

This flawed logic can burden minority groups with testing the performance of putatively fair---but actually misaligned---AI systems. Unfortunately, the emphasis on binary comparisons in fairness research creates further issues by implicitly positioning groups against each other, rather than collaborators against a system of oppression (cf. the ``racial wedge'' in political discourse;~\citealp{puri2016sexual}; and the trans-exclusionary movement in modern feminism;~\citealp{caslin2024trans, fahs2024urgent, serano2016whipping}). This combination of inequitable burdens and manufactured divisions ultimately undermines the very goals of algorithmic fairness, hindering the development of truly equitable and inclusive systems.

The binary approach to fairness not only oversimplifies complex social identities, but also tends to treat those identities as fixed and unchanging. Papers in our sample rarely discussed how or disclosed whether identity labels could change over time. By framing sensitive attributes as static and binary categories, algorithmic fairness research overlooks the dynamic and fluid nature of identity~\citep{keyes2019counting, tomasev2021fairness, lu2022subverting}. The fluidity and complexity of identity demands that algorithmic fairness research move beyond static and binary solutions, embracing approaches that can adapt to the intersectional and evolving nature of social inequalities.

\section{Discussion}
\label{sec:discussion}

Our meta-analysis surfaces two prominent trends in contemporary algorithmic fairness research: a disproportionate US-centric bias in author backgrounds, data origins, and design choices; and a widespread policy of casting sensitive attributes into binaries.

Despite popular depictions of research as an objective pursuit of truth, we recognize research as \textit{social praxis}---a process inherently shaped by contemporary norms and historical context~\citep{hochstein2019metaphysical, kuhn1961function, hacking1983representing}. Early in the development of a field, certain norms and design choices can help scope feasible research questions, providing an important starting point for researchers and scientists. However, over time, these initial frameworks become so deeply ingrained that they limit the scope of inquiry and exclude alternative perspectives.

US-centrism and binary logic represent two such frameworks. Current praxis advances a myth of algorithmic fairness as a fungible, abstract, and universal phenomena. However, this myth breaks apart when confronted with the dynamism of identity and the plurality of experience around the world.
To move beyond this myth, the fairness research community must chart new paths forward, guided by the principles of inclusivity, representativeness, and cultural specificity. 

The AIES and FAccT conferences reflect two important sites of dialogue for the algorithmic fairness community field. As a result, this meta-analysis offers a valuable snapshot of current trends in fairness research. In the future, expanding the scope of inquiry to other sites of study will help enrich and contextualize our understanding of contemporary norms and practices in fairness research. Promising sites include generalist conference venues (e.g., the Conference and Workshop on Neural Information Processing Systems), pre-print repositories (e.g., arXiv), and journals (e.g., \textit{Big Data \& Society}). Future work should also extend beyond the examination of academic discourse. An important step for this line of research will be to investigate how developers conceptualize and implement fairness within the systems they deploy to the real world. A critical examination of the social and structural factors shaping research practices across these sites will help identify biases and encourage a more inclusive and representative vision of algorithmic fairness.

Researchers and ethicists have proposed several solutions in this direction. Many prioritize transparency and self-reflection, advocating for explicit disclosure of the values embedded in technical work and operational definitions (e.g.~\citealp{van2023methodology}). Others emphasize diversifying representation across datasets (e.g.~\citealp{sambasivan2021seeing}), within research teams (e.g.~\citealp{laufer2022four}), and through participatory initiatives (e.g.~\citealp{gadiraju2023wouldn}). A final set of solutions focuses on grounding research, calling for greater engagement with real-world issues (e.g.~\citealp{birhane2022forgotten}) and rigorous ethnographic research (e.g.~\citealp{hagerty2019global, marda2021importance, martin2020extending}). %

The current focus on the US and on binary representations of sensitive attributes threatens to perpetuate a narrow and misleading perspective on algorithmic fairness. To overcome these limitations, the research community must expand its scope to encompass diverse social and cultural settings, particularly those outside the Global North. This expansion also requires moving beyond simplistic binary classifications and engaging with the involute, intersectional realities of bias, identity, and community~\citep{selbst2019fairness}. Qualitative methods and participatory approaches~\citep{birhane2022power, martin2020extending} will be crucial for developing datasets and fairness frameworks tailored to these conditions. Overall, researchers should seek to challenge the homogenizing effect of purely technical solutions and recenter the complex lived experiences of marginalized communities.

\section{Conclusion}
Our meta-analysis of algorithmic fairness papers reveals a field grappling with several fundamental tensions. While researchers strive to develop fair and unbiased systems, current research practices often default to US-centric perspectives and binary representations of sensitive attributes. As a result, current praxis not only can compromise the credibility of research insights, but also may lead to policy solutions that are poorly aligned with the needs of diverse communities. Our findings underscore the urgent need for fairness researchers to reflect on the norms embedded in their work, echoing and expanding calls for more inclusive, representative, and context-specific approaches to algorithmic fairness.

\section{Positionality Statement}

This paper emerges from a collective effort by researchers situated at the intersection of technology and marginalized communities. Our research backgrounds include the fields of algorithmic fairness, human-computer interaction, and critical algorithmic studies, with prior experience studying communities outside the Global North as well as issues afflicting LGBTQ+ populations. These experiences motivate our critical perspective on the existing AI ethics landscape and its inherent biases towards the US and the Global North. The lead author, for example, is a US-based doctoral student holding multiple intersecting identities—some marginalized, and others affording greater privilege. Navigating these complexities shapes their perspective on power dynamics within AI ethics discourse. Thus, we draw on our own identities and experiences to understand and motivate our research. While this personal perspective provides valuable insights, we acknowledge that it also introduces potential biases and welcome constructive engagement and alternative perspectives to challenge and enrich out understanding.

\section{Acknowledgements}
We would like to thank Seliem El-Sayed, Shakir Mohamed, and Simon Osindero for reviewing this work. We would also like to thank Iason Gabriel for his comments, insights, and advice throughout the process.

\bibliography{ref}

\clearpage

\appendix

\section{Expanded Results}
\label{App::Additional Findings}

\subsection{Data and Authorship Provenance}
\label{App::Authorship Affiliation By Year}
\begin{figure}[htbp]
\centering
\includegraphics[trim={0cm 3cm 0cm 0.5cm},width=0.75\textwidth]{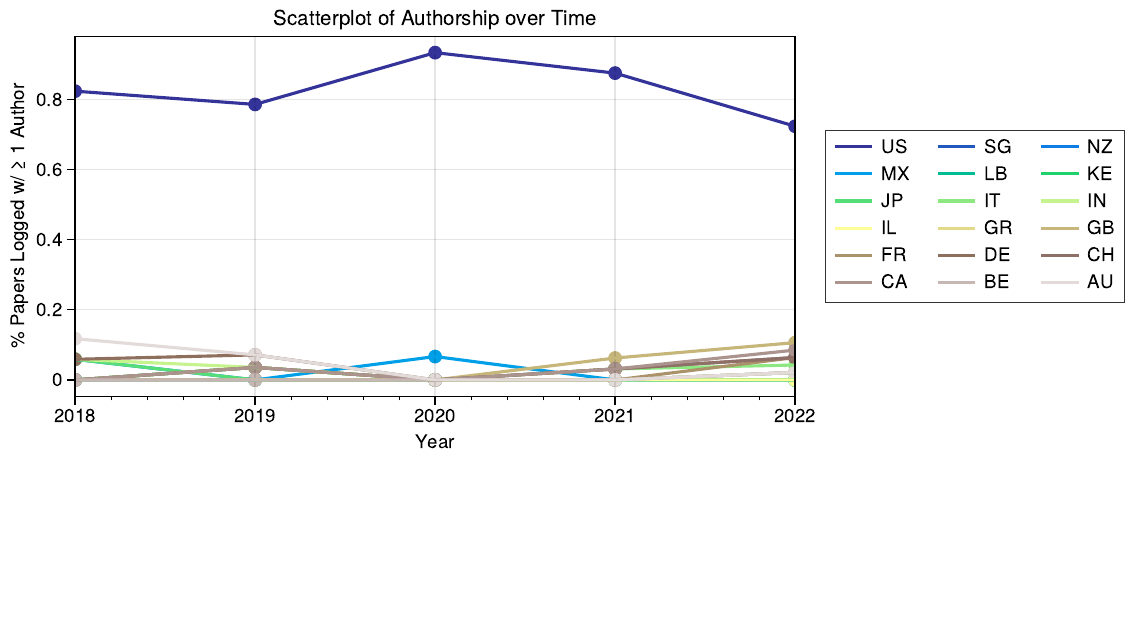}
\caption{Geographic Representation in Authorship Affiliations Over Time. The US consistently contributes the largest share of affiliations over all years included in our sample.}
\label{Fig::Authorship Affiliation By Year}
\end{figure}

\hyperref[Fig::Authorship Affiliation By Year]{Figure~\ref{Fig::Authorship Affiliation By Year}} shows the geographic distribution of authorship affiliations over time. \vspace{2em}

\subsection{Study Design}
\label{App::Study Design}
\begin{figure}[htbp]
\centering
\includegraphics[trim={0.5cm 5.9cm 0.5cm 0.45cm},width=0.95\textwidth]{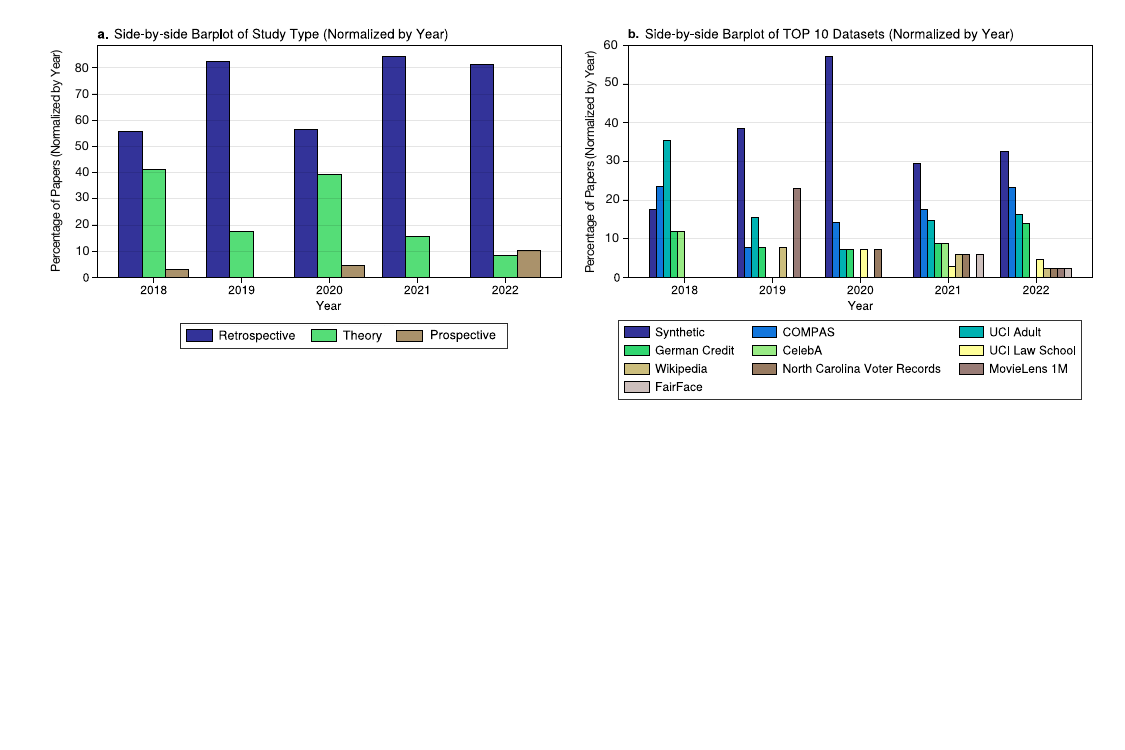}
\caption{Prevalence of Study Types and Datasets Over Time. (a) The majority of fairness papers engage in retrospective analysis of existing datasets and algorithms. (b) Studies most frequently leverage synthetic datasets. The most studied individual dataset is COMPAS, with other non-synthetic datasets like the German Credit and UCI Adult datasets also seeing considerable use.}
\label{Fig::Study Design}
\end{figure}

Fairness research papers can be categorized by the methods they apply to their proposed fairness definitions: empirical experiments that collect new data with a deployed intervention (\textit{prospective}); empirical simulations with existing datasets (\textit{retrospective}); or proofs using mathematical guarantees (\textit{theory}). Examining our sample through the lens of these categories reveals a clear trend: papers most frequently rely on retrospective analyses to study algorithmic fairness (\hyperref[Fig::Study Design]{Figure~\ref{Fig::Study Design}}a). This approach predominates across all years in our sample. In terms of dataset type, we observe a notable shift towards using synthetic data after 2018. The most common non-synthetic datasets (e.g., COMPAS, UCI Adult, and German Credit) disclose a focus on applications within financial and criminal justice contexts, though the exact pattern varies from year to year  (\hyperref[Fig::Study Design]{Figure~\ref{Fig::Study Design}}b).

\begin{figure}[htbp]
\centering
\includegraphics[trim={0cm 6cm 0cm 0.5cm},width=0.95\textwidth]{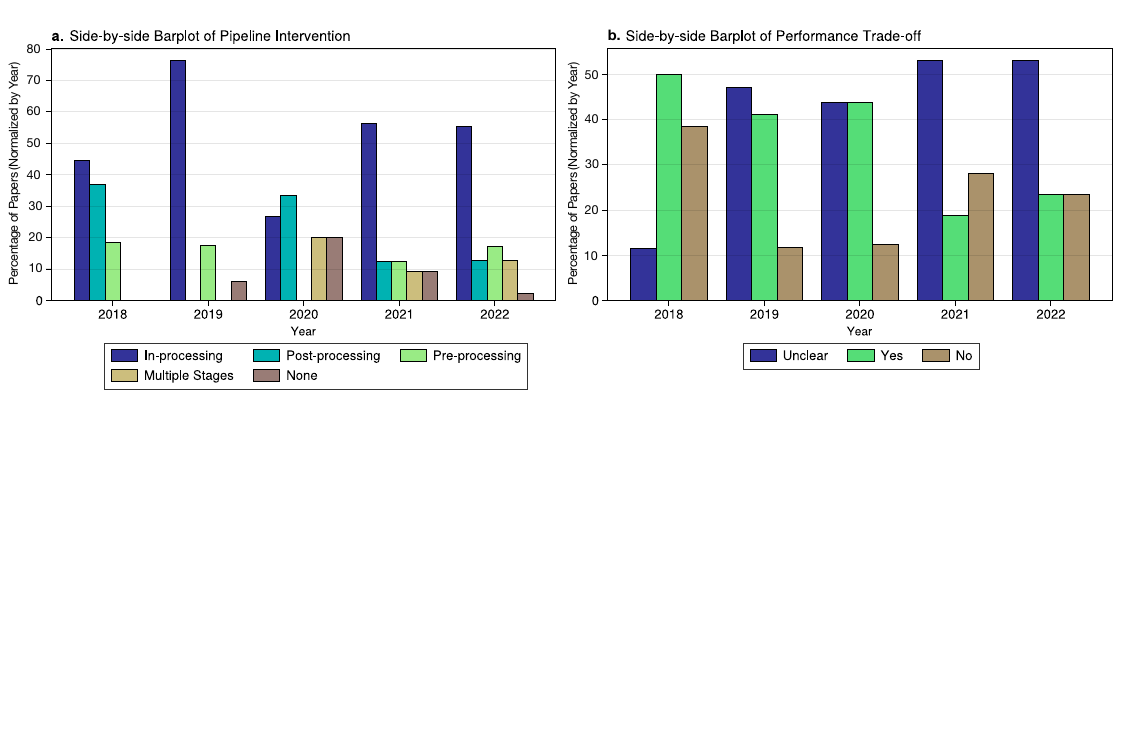}
\caption{Trends in Intervention Points and Acknowledgement of Performance Trade-offs Over Time. (a) In-processing consistently emerges as the most common point for implementing fairness interventions, with the exception of 2020. (b) Notably, we observe an upward trend in the number of papers that do not address whether improving fairness metrics comes at the cost of performance.}
\label{Fig::Bigger Picture}
\end{figure}

We next analyze aspects of the algorithmic interventions proposed. Our results indicate a clear preference for in-processing interventions, which directly modify the algorithm during the learning process (\hyperref[Fig::Bigger Picture]{Figure~\ref{Fig::Bigger Picture}}a). In-processing approaches tend to overshadow pre-processing interventions, which focus on data adjustments before training, and post-processing interventions, which address fairness concerns by modifying algorithmic outputs. Across the years we examine, we observe a growing ambiguity regarding the inherent trade-off between fairness and performance (\hyperref[Fig::Bigger Picture]{Figure~\ref{Fig::Bigger Picture}}b).

\clearpage
\section{Fairness Definitions}
\label{App::Fairness Definitions}

\begin{figure}[htbp]
\centering
\includegraphics[trim={2cm 2.5cm 2cm 0.1cm},width=0.75\textwidth]{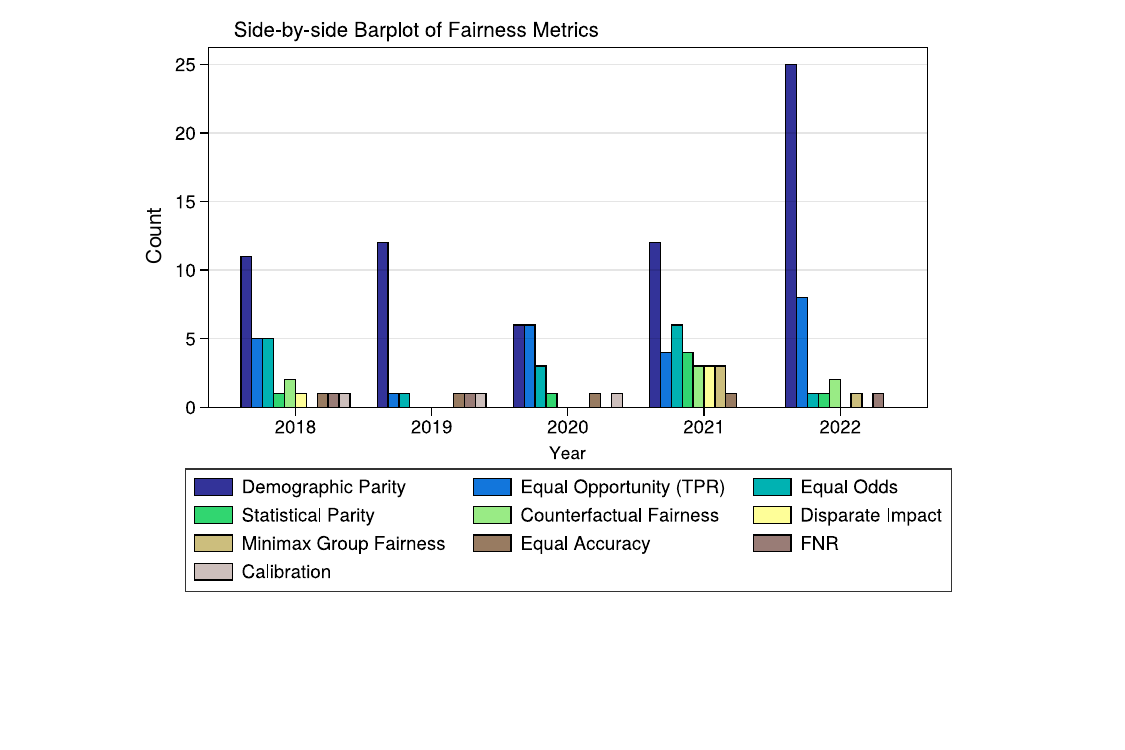}
\caption{Popularity of Top Fairness Metrics Over Time. While a wide range of metrics exist for assessing fairness, a small subset sees disproportionate use within the literature. This barplot depicts the ten most common fairness metrics from our sample. Demographic parity demonstrates consistent popularity over the years.}
\label{Fig::Metric Names Over Time}
\end{figure}

\begin{figure}[htbp]
\centering
\includegraphics[width=0.95\textwidth]{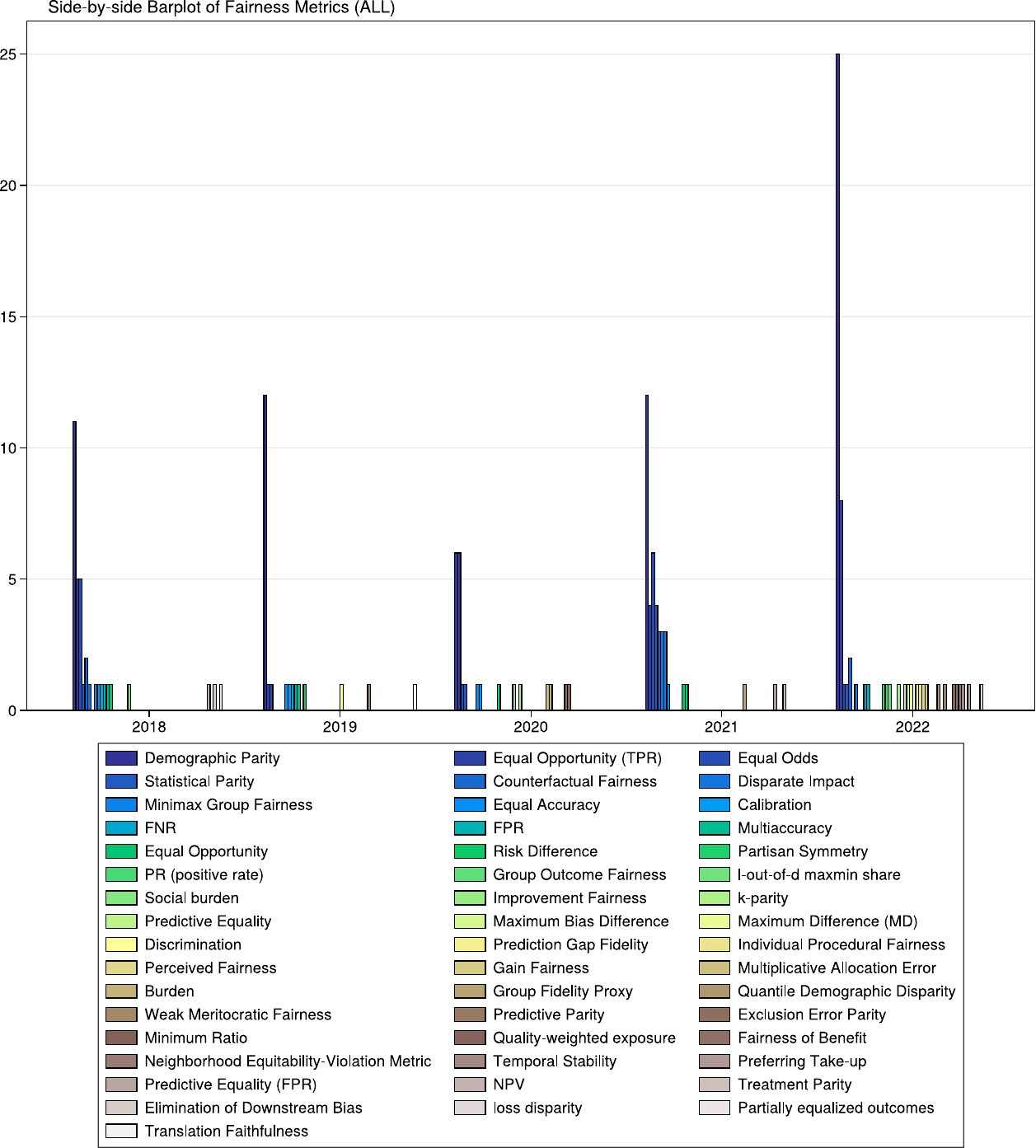}
\caption{Prevalence of All Fairness Metrics Over Time. This barplot depicts the use of fairness metrics across papers in our sample.}
\label{Fig::Metric Names Over Time All}
\end{figure}

Figures~\ref{Fig::Metric Names Over Time} and~\ref{Fig::Metric Names Over Time All} illustrate the prevalence of different fairness metrics in our sample over time, presenting the top ten most frequently used metrics and the full list, respectively. A single study might explore multiple metrics. For instance, an individual paper might examine both demographic parity and equalized false positive rates across demographic groups.

\label{App::Fairness Definitions All}

\newpage

\section{Sampled Papers}
\label{App::Included Papers}
Tables~\ref{Table::Included_Papers_1} and~\ref{Table::Included_Papers_2} 
provide titles for the full sample of papers selected and reviewed for our meta-analysis.

\begin{table}[]
    \centering
    \includegraphics[width=0.95\textwidth]{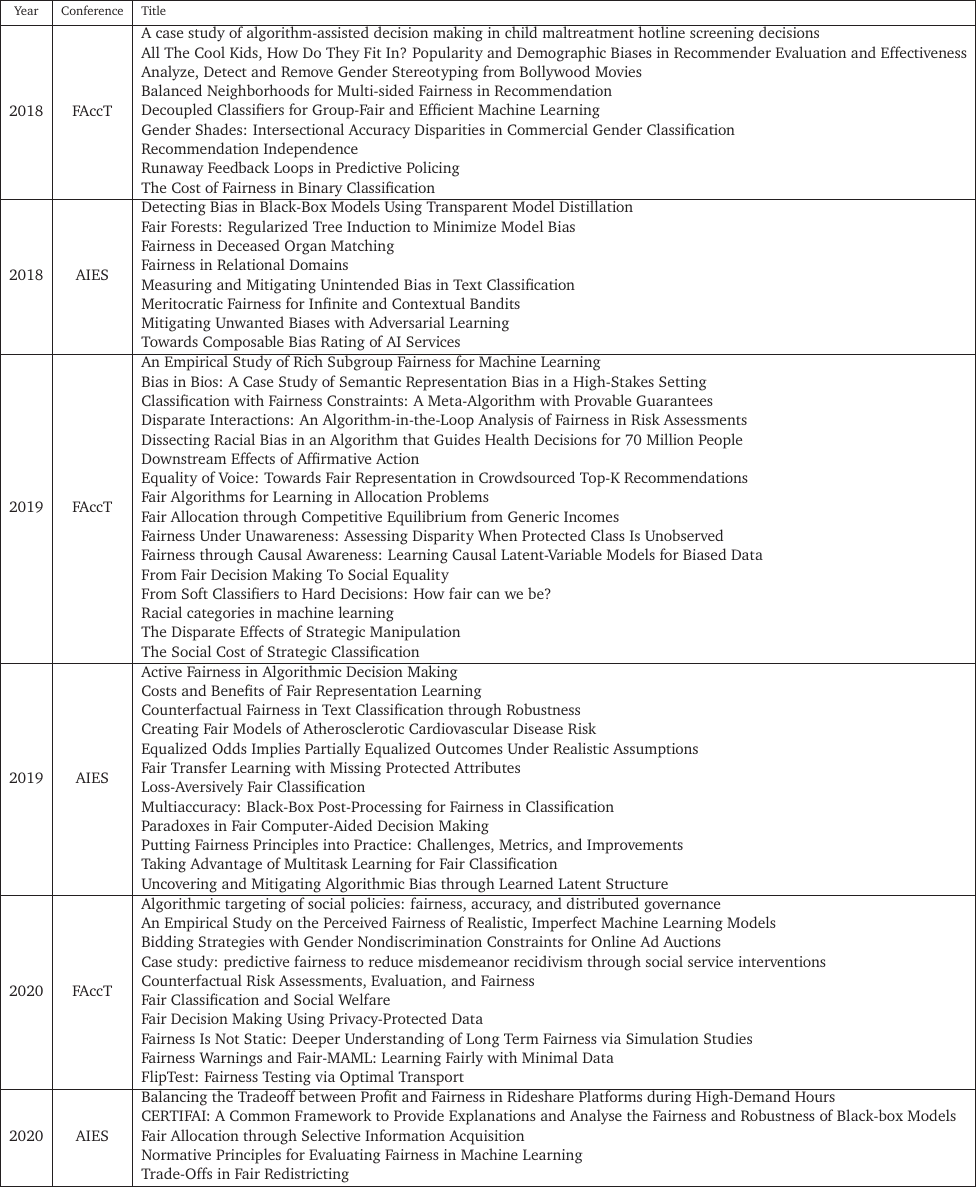}
\caption{Sampled Papers, from 2018 to 2020.}
\label{Table::Included_Papers_1}
\end{table}

\begin{table}[]
    \centering
    \includegraphics[width=0.95\textwidth]{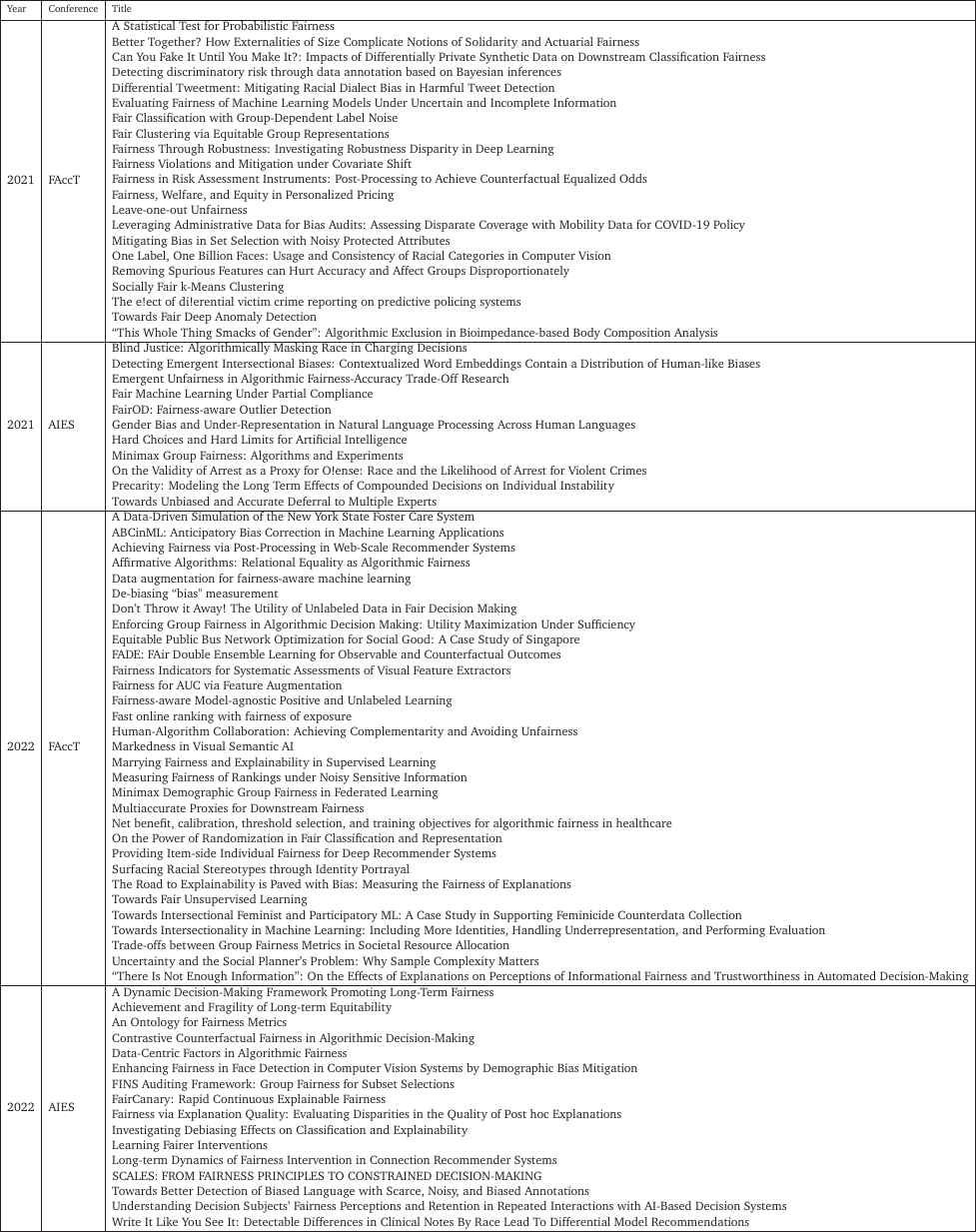}
\caption{Sampled Papers, from 2021 to 2022.}
\label{Table::Included_Papers_2}
\end{table}

\end{document}